\documentclass[prd,onecolumn]{revtex4}
\usepackage{dcolumn}
\usepackage{multirow}
\usepackage{graphicx}
\usepackage{amssymb}
\usepackage{bm}
\usepackage{hyperref}%for .pdf
\usepackage{epstopdf}%for .pdf
\usepackage{color}
\usepackage{mathrsfs}
\usepackage{amsmath,amssymb,amsthm}
\begin{document}
\title{Constraining viscous dark energy models with the latest cosmological data}

\author{Deng Wang$^{1}$}
\email{Cstar@mail.nankai.edu.cn}
\author{Yang-Jie Yan$^{2}$}
\email{yanyj926@gmail.com}
\author{Xin-He Meng$^{2}$}
\email{xhm@nankai.edu.cn}

\affiliation{
$^1${Theoretical Physics Division, Chern Institute of Mathematics, Nankai University, Tianjin 300071, China}\\
$^2${Department of Physics, Nankai University, Tianjin 300071, China}\\}
\begin{abstract}
Based on the assumption that the dark energy possessing bulk viscosity is homogenously and isotropically permeated in the universe, we propose three new viscous dark energy (VDE) models to characterize the accelerating universe. By constraining these three models with the latest cosmological observations, we find that they just deviate very slightly from the standard cosmological model and can alleviate effectively the current $H_0$ tension between the local observation by the Hubble Space Telescope and the global measurement by the Planck Satellite. Interestingly, we conclude that a spatially flat universe in our VDE model with cosmic curvature is still supported by current data, and the scale invariant primordial power spectrum is strongly excluded at least at the $5.5\sigma$ confidence level in three VDE models as the Planck result. We also give the $95\%$ upper limits of the typical bulk viscosity parameter $\eta$ in three VDE scenarios.

\end{abstract}
\maketitle
\section{Introduction}
There is no doubt that modern cosmology has entered a new data-driven era. During the past nearly two decades, a large number of observational evidences such as Type Ia supernovae (SNIa) \cite{1,2}, baryonic acoustic oscillations (BAO) \cite{3} and cosmic microwave background (CMB) anisotropies \cite{4,5} have confirmed the late-time acceleration of the universe and verified that our universe consists approximately of $30\%$ matter including baryons and cold dark matter (CDM) and $70\%$ dark energy components. The simplest candidate of dark energy is the so-called cosmological constant scenario, namely the $\Lambda$CDM model, which has been confirmed as a phenomenologically viable cosmological model, once again, by the recent Planck CMB measurements with high precision \cite{6}. However, this well-known model suffers from two severe problems \cite{7}: (i) Why is the value of the cosmological constant unexpectedly small with respect to
any physically meaningful scale, except the current horizon scale ? (ii) Why this value is not only small, but also surprisingly close to another unrelated physical quantity, the present-day matter density ? Up to date, due to the fact that we still know little about the realistic nature of the dark energy, to resolve these puzzles, a flood of candidates have been proposed by cosmologists. Generally, they can be divided into two main classes, i.e., the dark energy models \cite{8,9,10,11,12,13,14,15,16,17,18,19,20,21,22,23,24} and modified theories of gravity (GR) \cite{25,26,27,28,29,30,31,32}. The former case is aimed at introducing a new cosmic fluid or matter field within the framework of the theory of general relativity. On the contrary, assuming the breakdown of Einstein's gravity, the latter case modifies the standard GR lagrangian based on some concrete physical mechanism.

Based on the thermodynamics point of view, dissipative process is a universal property of any realistically physical phenomena. In the literature, as a consequence, several authors \cite{33,34,35} have naturally used the exotic fluids from dissipative process, which belongs to the former case mentioned above, to explain the cosmic acceleration. The concretely physical realization of dissipative process is the addition of shear and bulk viscosity terms into the stress-energy tensor of a cosmic fluid \cite{36,37,38,39,40,41,42,43}. Furthermore, assuming the cosmological principle --- the isotropic and homogenous universe, the shear viscosity vanishes and one only needs to consider the impacts of bulk viscosity on the evolution of the universe in a given cosmological model. The bulk viscosity effects have been widely investigated in the cosmological context, for example, inflation in a viscous fluid model \cite{44}, the gravitino abundance within bulk viscosity cosmology \cite{45}, viscous dark matter (VDM) \cite{46,47,48,49}, viscous dark energy (VDE) \cite{14,15,16,17,18,19,50,51,52,53,54,55}, and so forth. Several sources of viscous fluid can be described as follows: moving cosmic strings through the comic magnetic fields, magnetic monopoles in monopole interactions effectively experience various bulk viscosity phenomena \cite{56,57} or primordial quantum particle productions and their interactions \cite{58,59,60,61,62,63}.

By observing the Einstein equations, one can find that the left-hand side involves the geometry or gravity, and on the other side the stress-energy tensor is also related to a quantum field vacuum. Starting directly from the equation of state (EoS), VDE models are aimed at studying the evolutionary properties of the universe and explore whether a given EoS of DE can effectively characterize the observational data. A majority of previous VDE models are dedicated to single component for the dark sector, i.e., regarding the DM and DE as a unified dark fluid \cite{14,15,16,17,18,19,23,24,64}. Interestingly, the correspondences between the modified EoS of viscous cosmological models, the scalar fields, and extended theories of gravity are also investigated in details \cite{15,65,66,67}. It is noteworthy that the VDE models have a convenient way to resolve the so-called Big Rip problem \cite{68,69}, which is encountered by the phantom cosmology \cite{13}, since their EoS can easily exhibit a phantom-barrier-crossing behavior via the bulk viscosity terms (see \cite{v} for a recent review).

In the present work, we focus on proposing three new VDE models and use the latest cosmological observations including Planck data to constrain them. In the meanwhile, we also study the abilities of these three models in alleviating the current Hubble constant $H_0$ tension between the indirectly global measurement by the Planck satellite \cite{70} and the directly local measurement by the Hubble space telescope \cite{71}. We find that they just deviate very slightly from the $\Lambda$CDM model and can relieve effectively the $H_0$ tension.

This work is organized as follows. In the next section, we introduce briefly our three new VDE models. In Section III, we investigate numerically the effects of bulk viscosity on the evolutional behaviors of the universe in these models.
 In Section IV, we describe the latest observational data and our analysis method. In Section V, we exhibit our results. The discussions and conclusions are presented in the final section.

\section{VDE models}
The dynamics of a Friedmann-Robertson-Walker (FRW) universe are governed by the following two Friedmann equations
\begin{equation}
\frac{\dot{a}^2}{a^2}=\frac{\rho}{3}, \label{1}
\end{equation}
\begin{equation}
\frac{\ddot{a}}{a}=-\frac{\rho+p}{6}, \label{2}
\end{equation}
where $a$, $\rho$ and $p$ are the scale factor, energy density and pressure of the cosmic fluid, respectively, and the dot denotes the derivative with respect to the cosmic time $t$. Note that we use the units $8\pi G=c=\hbar=1$ throughout this work. Subsequently, we consider a cosmic DE fluid possessing the bulk viscosity $\zeta$ and its stress-energy tensor is written as
\begin{equation}
T_{\mu\nu}^{de}=\rho_{de} U_{\mu}U_{\nu}+\tilde{p}_{de}h_{\mu\nu}, \label{3}
\end{equation}
where $\rho_{de}$ and $\tilde{p}_{de}$ are the energy density and effective pressure of VDE fluid, $U_{\mu}=(1, 0, 0, 0)$ the four-velocity of the VDE fluid in comoving coordinates and $h_{\mu\nu}=g_{\mu\nu}+ U_{\mu}U_{\nu}$ the projection tensor. For a typical VDE fluid, according to Ekart's theory as a first order limit of Israel-Stewart scenario with zero relation time, one can re-express the effective pressure $\tilde{p_{de}}$ at the thermodynamical equilibrium as $\tilde{p_{de}}=\omega\rho_{de}-\zeta\theta$ [], where $\omega$, $\zeta$ and $\theta$ is the EoS of perfect DE fluid, bulk viscosity and expansion scalar, respectively. Furthermore, the corresponding energy conservation equation for the VDE fluid can be written as
\begin{equation}
\dot{\rho_{de}}+\theta(\rho_{de}+\tilde{p_{de}})=0, \label{4}
\end{equation}
where $\theta=3H=3\frac{\dot{a}}{a}$ and $H$ is the Hubble parameter that controls the background evolution of the universe.

We consider the bulk viscosity being proportional to the Hubble parameter, i.e., $\zeta=\eta H$ in our analysis from first to last. Consequently, the effective pressure reads $\tilde{p}_{de}=\omega\rho_{de}-3\eta H^2$ now.
For the first VDE model, we just consider a one-parameter extension to the standard six-parameter $\Lambda$CDM cosmology by fixing $\omega=-1$ and only adding the viscous terms $-3\eta H^2$ into the effective pressure. Hereafter we call it the V$\Lambda$DE model and its effective pressure is $\tilde{p_{de}}=-\rho_{de}-3\eta H^2$. Subsequently, combining Eqs. (\ref{1}) with (\ref{4}), the dimensionless Hubble parameter $E(z)$ of the V$\Lambda$DE model can be expressed as
\begin{equation}
E(z)=\left[\frac{1}{1+\eta}\Omega_{m}(1+z)^3+(1-\frac{1}{1+\eta}\Omega_{m})(1+z)^{-3\eta}\right]^{\frac{1}{2}},   \label{5}
\end{equation}
where $z$ and $\Omega_{m}$ are the redshift and present-day matter density parameter, respectively. Notice that for the V$\Lambda$DE model, we have ignored the contribution from the radiation and spatial curvature components. One can also find that the V$\Lambda$DE model reduces to the standard $\Lambda$CDM one when $\eta=0$ in Eq. (\ref{5}). For the second VDE model, we make a simple one-parameter generalization to the V$\Lambda$DE model by setting the EoS of DE $\omega$ as a free parameter and we call it the V$\omega$DE model hereafter. Furthermore, we can exhibit its dimensionless Hubble parameter $E(z)$ as
\begin{equation}
E(z)=\left[\frac{\omega}{\omega-\eta}\Omega_{m}(1+z)^3+(1-\frac{\omega}{\omega-\eta}\Omega_{m})(1+z)^{3(1+\omega-\eta)}\right]^{1/2}.   \label{6}
\end{equation}
For the third VDE model, for the first time, we consider the bulk viscosity effects on the spatial curvature by adding the curvature contribution into the cosmic pie. More specifically, we still consider a simple one-parameter extension to the V$\Lambda$DE model by fixing $\omega=-1$ and we call it the VKDE model in the following context. The corresponding dimensionless Hubble parameter $E(z)$ of the VKDE model is
\begin{equation}
E(z)=\left[\frac{2}{2+3\eta}\Omega_{k}(1+z)^2+ \frac{1}{1+\eta}\Omega_{m}(1+z)^3+(1-\frac{2}{2+3\eta}\Omega_{k}-\frac{1}{1+\eta}\Omega_{m})(1+z)^{-3\eta}\right]^{1/2},   \label{7}
\end{equation}
where $\Omega_{k}$ is the present-day curvature density parameter. Note that the $\Lambda$CDM cosmology recovers in this case when $\eta=\Omega_{k}=0$. Furthermore, combining the expression of the effective pressure of VDE fluid $\tilde{p_{de}}=-\rho_{de}-3\eta H^2$ and Eq. (\ref{4}), we obtain, respectively, the effective EoS of VDE fluid for the V$\Lambda$DE, V$\omega$DE and VKDE models as
\begin{equation}
\omega_{de}(z)=-1-\eta-\frac{\eta\Omega_{m}(1+z)^3}{(1-\frac{1-\eta}{1+\eta}\Omega_{m})(1+z)^{-3\eta}-\frac{\eta}{1+\eta}\Omega_{m}(1+z)^3}, \label{8}
\end{equation}
\begin{equation}
\omega_{de}(z)=\omega-\eta-\frac{\eta\Omega_{m}(1+z)^3}{(1-\frac{\omega+\eta}{\omega-\eta}\Omega_{m})(1+z)^{3(1+\omega-\eta)}+\frac{\eta}{\omega-\eta}\Omega_{m}(1+z)^3}, \label{9}
\end{equation}
\begin{equation}
\omega_{de}(z)=-1-\eta-\frac{\eta[\Omega_{m}(1+z)^3+\Omega_{k}(1+z)^2]}{(1-\frac{2-3\eta}{2+3\eta}\Omega_{k}-\frac{1-\eta}{1+\eta}\Omega_{m})(1+z)^{-3\eta}-\frac{\eta}{1+\eta}\Omega_{m}(1+z)^3-\frac{3\eta}{2+3\eta}\Omega_{k}(1+z)^2}.
\end{equation}

\begin{figure}
\centering
\includegraphics[scale=0.4]{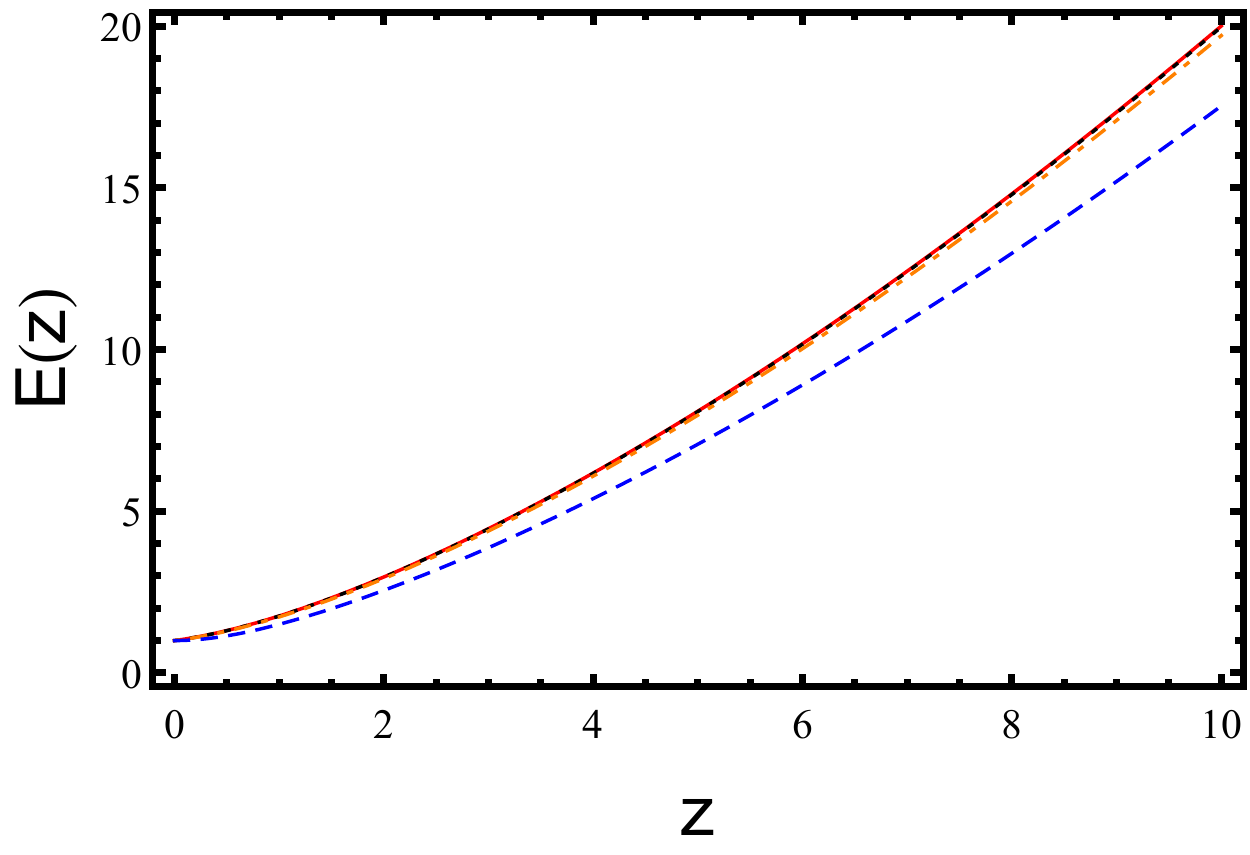}
\caption{The effects of the bulk viscosity coefficient $\eta$ on the dimensionless Hubble parameter $E(z)$ in the V$\Lambda$DE model. Here we have assumed $\Omega_{m}=0.3$ and the solid (red), dotted (black), dash-dotted (orange) and dashed (blue) lines correspond to the cases of $\eta=0$ ($\Lambda$CDM), 0.003, 0.03 and 0.3, respectively. }\label{f1}
\end{figure}

\begin{figure}
\centering
\includegraphics[scale=0.4]{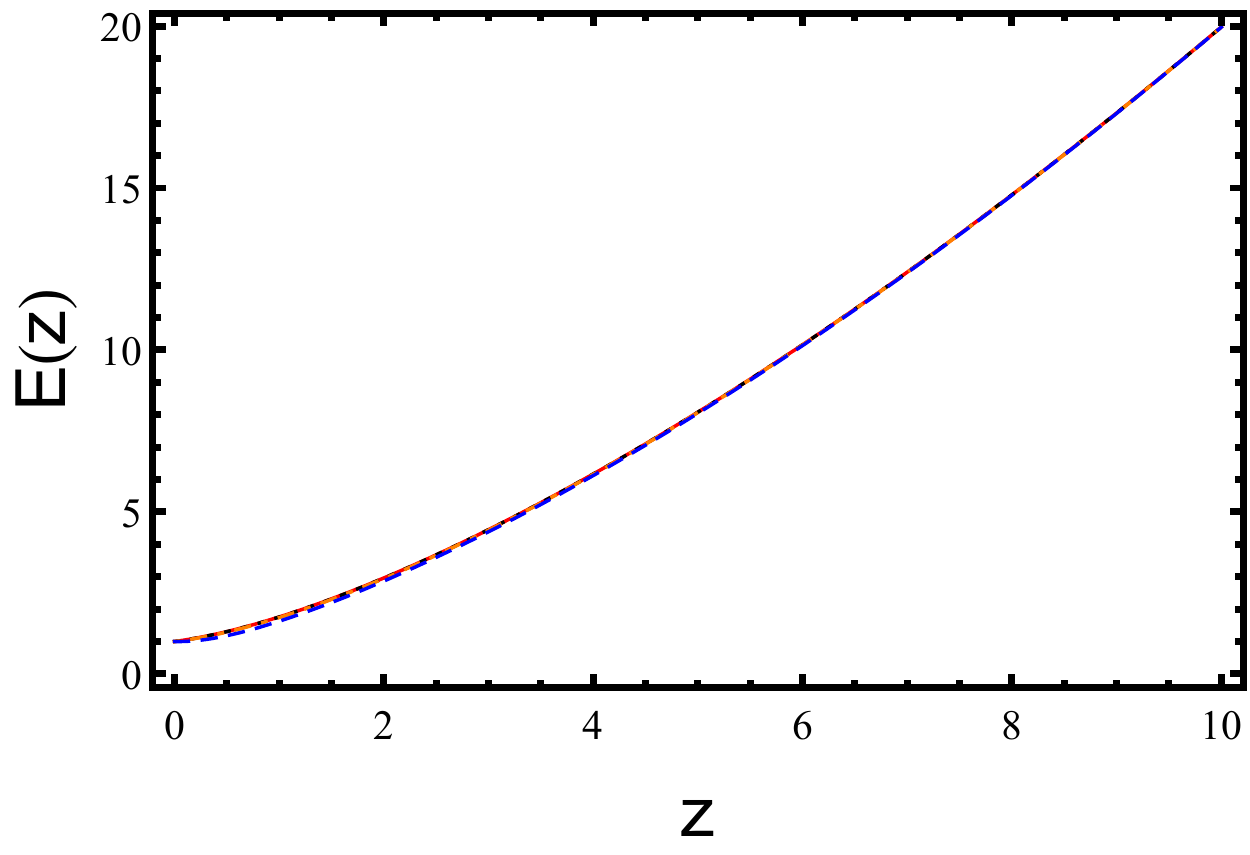}
\includegraphics[scale=0.4]{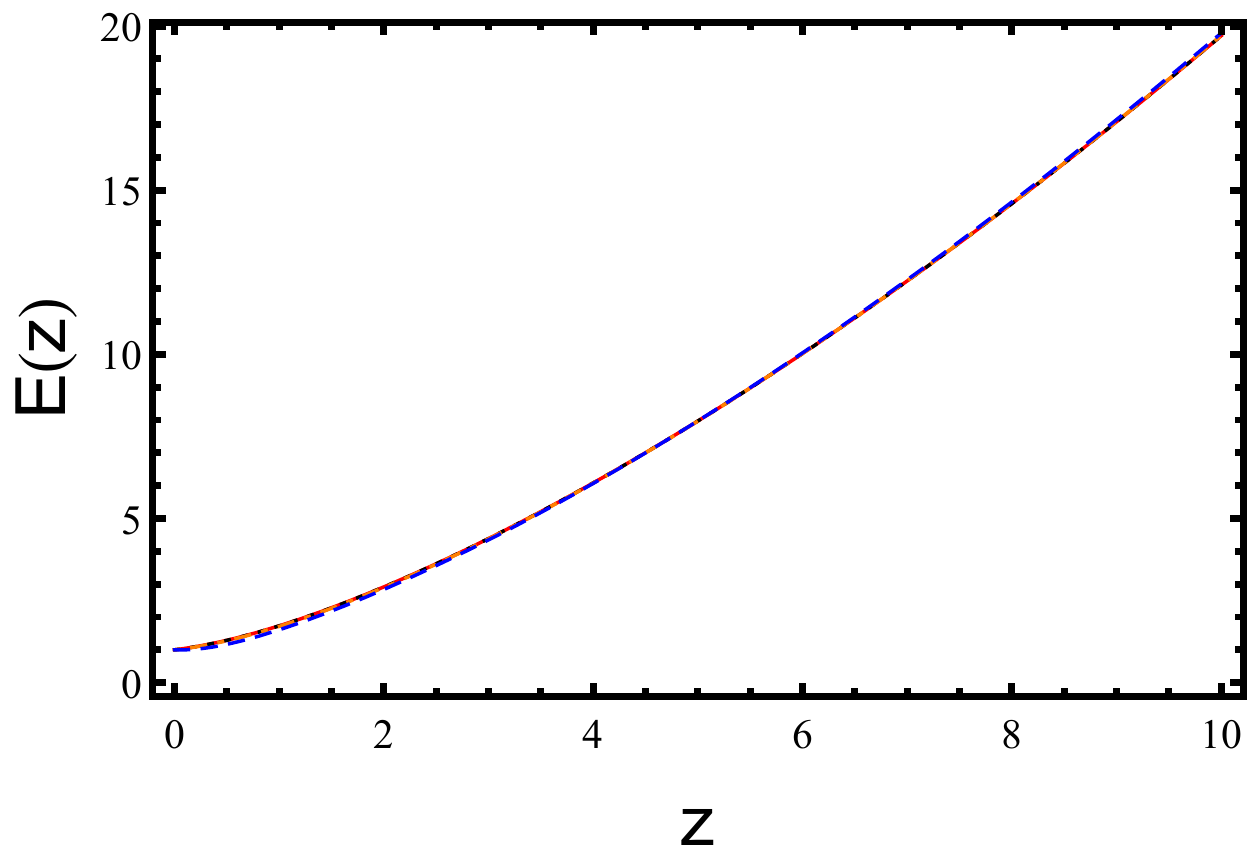}
\includegraphics[scale=0.4]{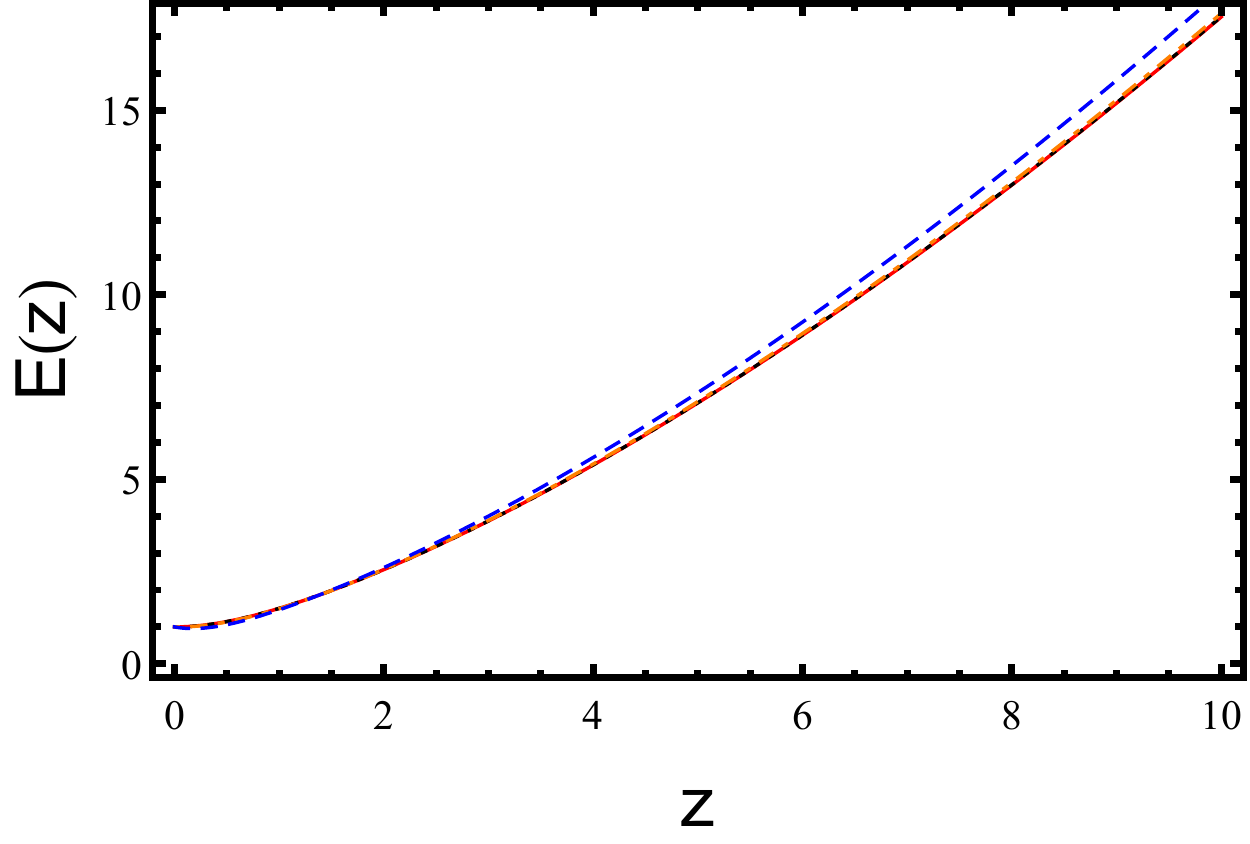}
\caption{The effects of the perfect EoS of DE $\omega$ on the dimensionless Hubble parameter $E(z)$ for different values of bulk viscosity coefficient $\eta$ in the V$\omega$DE model. From left to right, fixing $\Omega_{m}=0.3$, we consider the cases of $\eta=0.003, 0.03, 0.3$ and the solid (red), dotted (black), dash-dotted (orange) and dashed (blue) lines correspond to the case of $\omega=0$, -1.005, -1.05 and -1.5, respectively. }\label{f2}
\end{figure}

\begin{figure}
\centering
\includegraphics[scale=0.4]{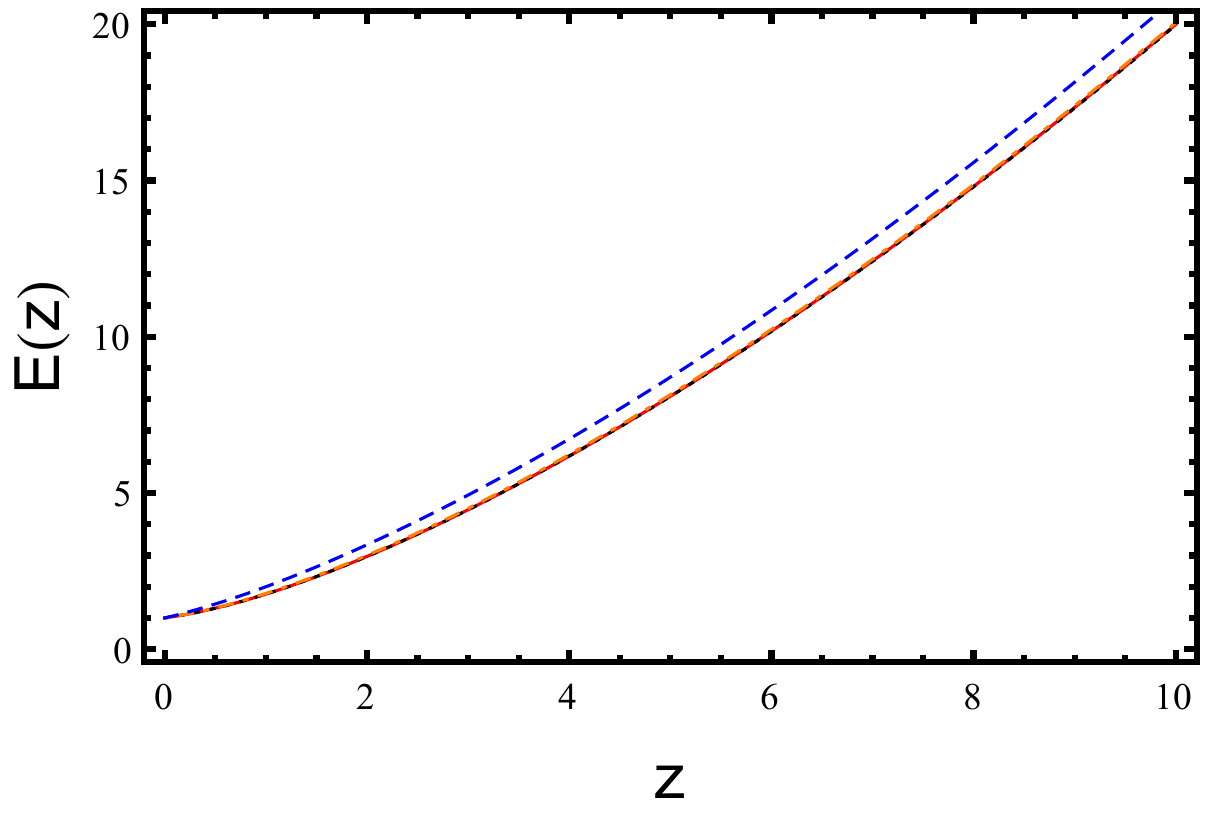}
\caption{The effects of the spatial curvature $\Omega_{k}$ on the dimensionless Hubble parameter $E(z)$ in the VKDE model. Here we have assumed $\Omega_{m}=0.3$ and $\eta=0.003$ and the solid (red), dotted (black), dash-dotted (orange) and dashed (blue) lines correspond to the cases of $\Omega_{k}=0$ ($\Lambda$CDM), 0.003, 0.03 and 0.3, respectively. }\label{f3}
\end{figure}

\begin{figure}
\centering
\includegraphics[scale=0.4]{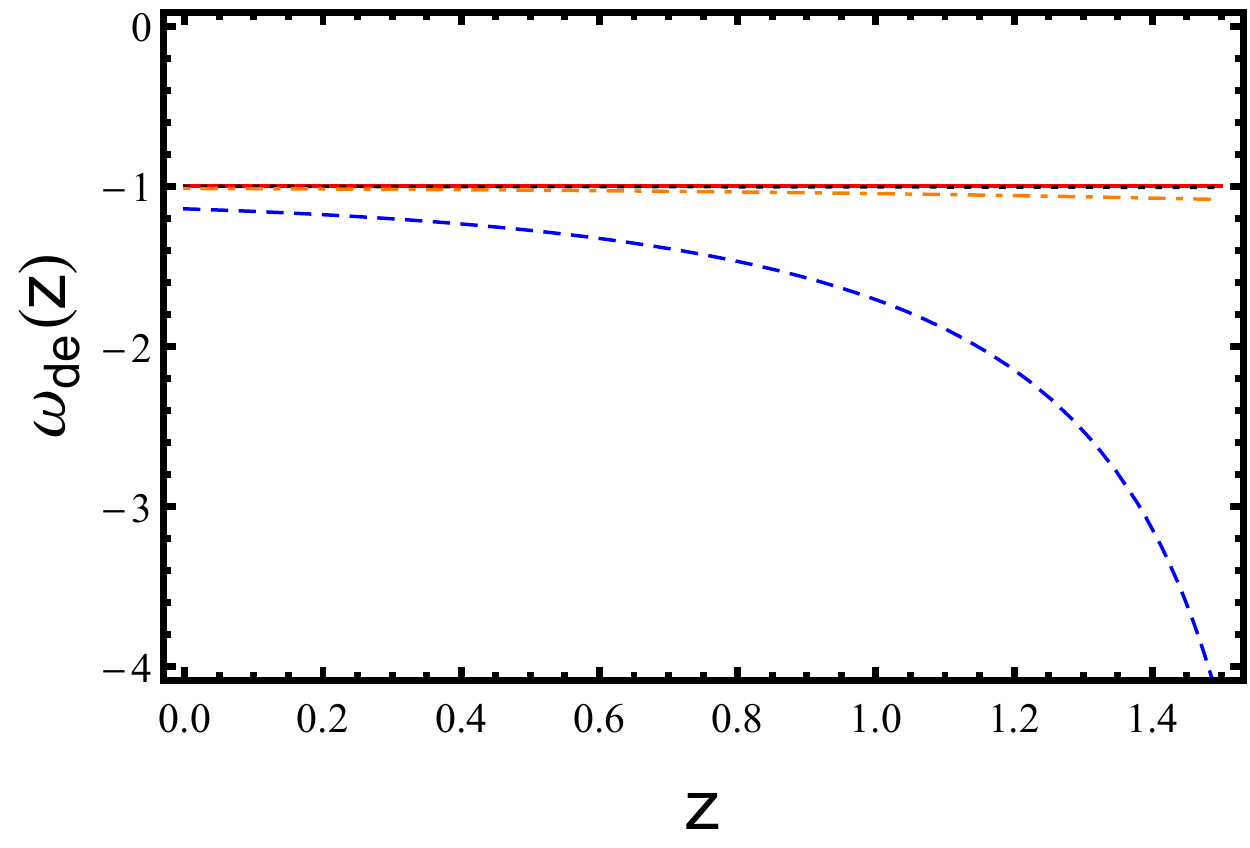}
\includegraphics[scale=0.4]{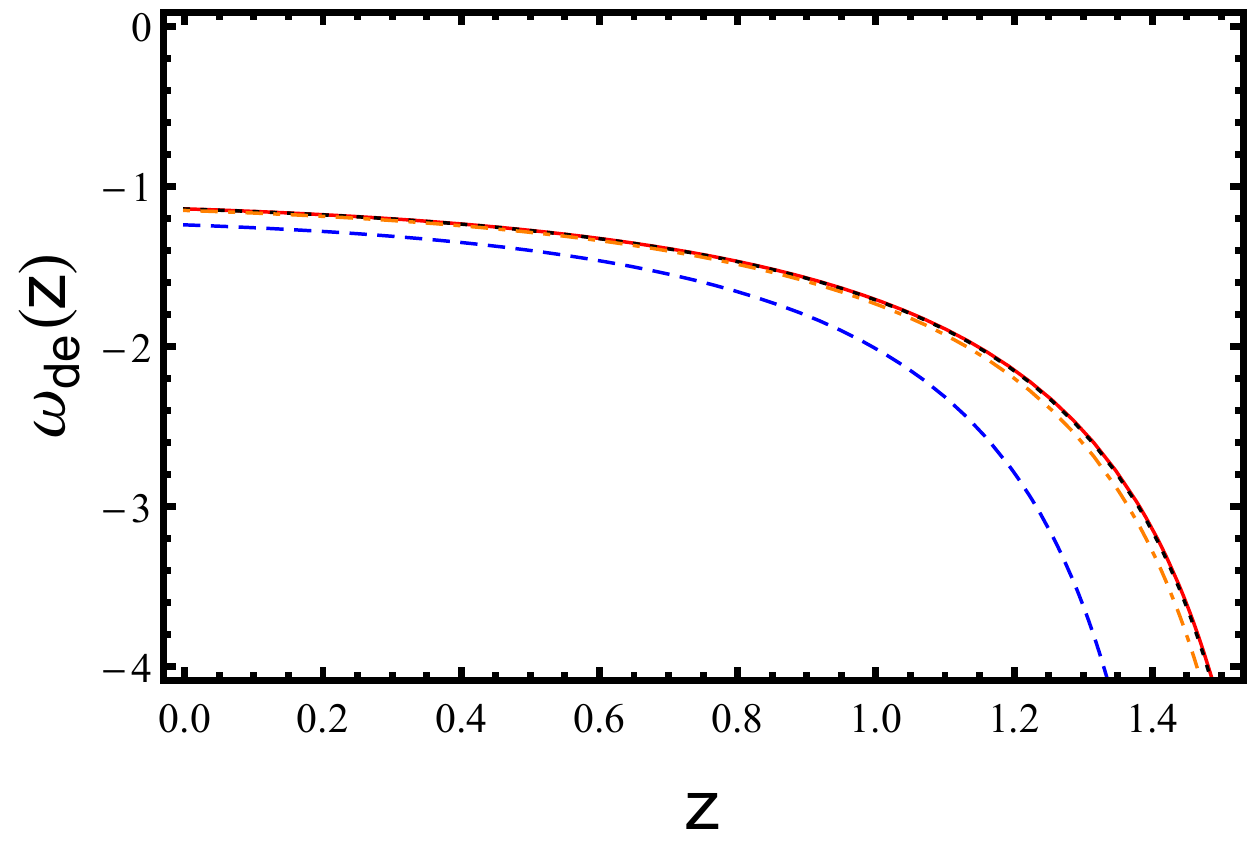}
\includegraphics[scale=0.44]{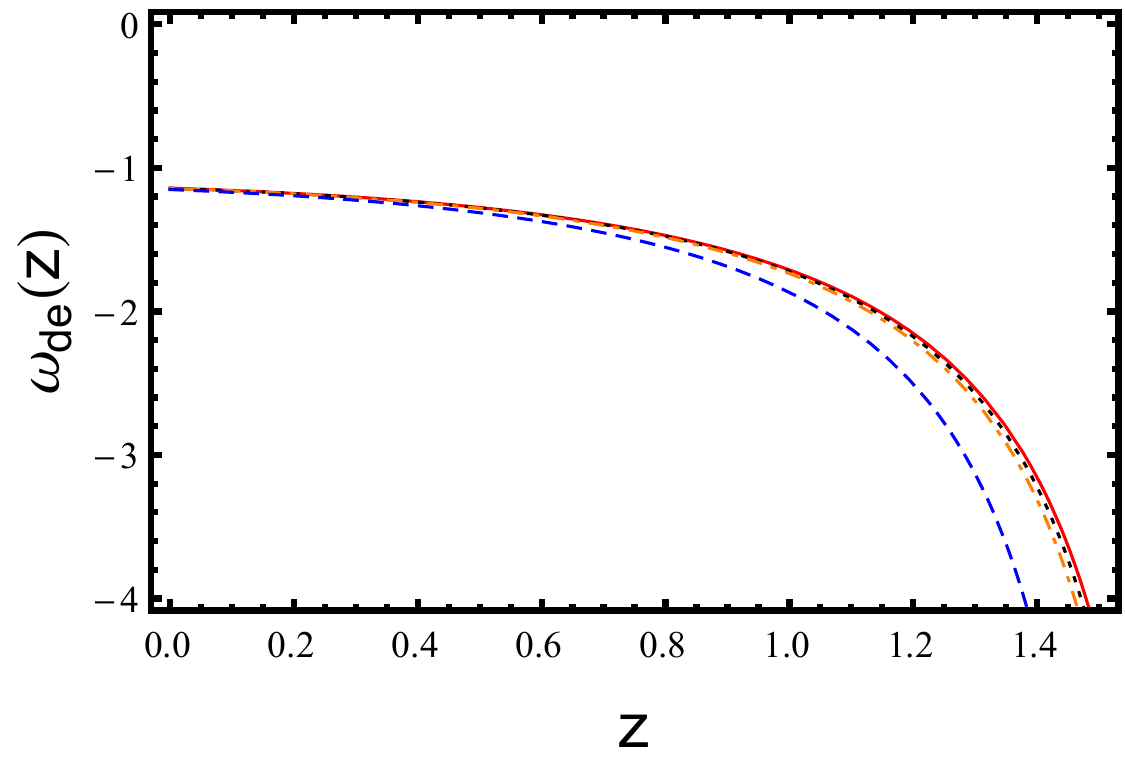}
\caption{\textit{Left panel}: The effects of the bulk viscosity coefficient $\eta$ on the effective EoS of VDE $\omega_{de}(z)$ in the V$\Lambda$DE model. The solid (red), dotted (black), dash-dotted (orange) and dashed (blue) lines correspond to the cases of $\eta=0$ ($\Lambda$CDM), 0.001, 0.01 and 0.1, respectively. \textit{Middle panel}: The effects of the perfect EoS of DE $\omega$ on the effective EoS of VDE $\omega_{de}(z)$ in the V$\omega$DE model. Fixing $\eta=0.1$, the solid (red), dotted (black), dash-dotted (orange) and dashed (blue) lines correspond to the cases of $\omega=-1$, -1.001, -1.01 and -1.1, respectively. \textit{Right panel}: The effects of the spatial curvature $\Omega_{k}$ on the effective EoS of VDE $\omega_{de}(z)$ in the VKDE model. Fixing $\eta=0.1$, the solid (red), dotted (black), dash-dotted (orange) and dashed (blue) lines correspond to the cases of $\Omega_{k}=0.001$, 0.005, 0.01 and 0.05, respectively. Here we have assumed $\Omega_{m}=0.3$.}\label{f4}
\end{figure}

\section{Qualitative analysis of bulk viscosity effects on the evolution of the universe}
It is very interesting to study the impacts of VDE permeated homogenously at cosmological scales on the evolution of the universe. To be more specific, we hope to obtain the qualitative relations between the bulk viscosity coefficient $\eta$ and two background physical quantities, i.e., the dimensionless Hubble parameter $E(z)$ and effective EoS of VDE $\omega_{de}(z)$, in our three VDE models.

From Fig. \ref{f1}, we find that the larger the bulk viscosity is, the smaller the dimensionless Hubble parameter is, i.e., the smaller the expansion rate of the universe is (see also Eq. (\ref{5})). However, the effects of bulk viscosity decrease evidently at low redshifts unlike at high ones, and the evolution of the universe in the V$\Lambda$DE model behaves very close to that of the standard cosmology. Note that this conclusion is not only valid for the V$\Lambda$DE model but also holds true for all the simple VDE models. Subsequently, we also study the impacts of the perfect freedom $\omega$ on the expansion rate of the universe in the V$\omega$DE model by fixing $\eta=0.003$, 0.03 and 0.3, respectively. From Fig. \ref{f2}, we obtain the conclusion that, for a universe where $\omega$ is negative enough (e.g. $\omega=-1.5$) and bulk viscosity is large enough (e.g. $\eta=0.3$), the more negative the perfect EoS of DE $\omega$ is, the faster the universe expands (see the right panel of Fig. \ref{f2}). Furthermore, if taking the spatial curvature into account, we find that the larger the present-day curvature density $\Omega_{k}$ is, the faster the universe expands (see Fig. \ref{f3}). However, this holds only for a very large spatial curvature. If considering the recent restriction $|\Omega_{k}|<0.005$ by the Planck Satellite \cite{6}, one can find that the variation of $\Omega_{k}$ affects hardly the evolution of the universe. For the VDE models, we are also very interested in the evolutional behaviors of the effective EoS of VDE. Form Fig. \ref{f4}, we find that a more negative effective EoS of VDE $\omega_{de}(z)$ corresponds to a larger bulk viscosity $\eta$ in the V$\Lambda$DE model, a more negative perfect EoS of DE $\omega$ in the V$\omega$DE model and a larger cosmic curvature $\Omega_{k}$ in the VKDE model, respectively. Moreover, our three VDE models are very close to the $\Lambda$CDM one when $z$ approaches zero.

The above-mentioned investigations are just some simple qualitative analysis about the evolutional behaviors of VDE models. In the next section, we utilize the latest cosmological observations to constrain these three VDE models and their corresponding parameter spaces are presented as follows:
\begin{equation}
\mathbf{P_{V\Lambda DE}}=\{\Omega_bh^2, \quad \Omega_ch^2, \quad 100\theta_{MC}, \quad \tau, \quad \eta, \quad  \mathrm{ln}(10^{10}A_s), \quad  n_s \},   \label{11}
\end{equation}
\begin{equation}
\mathbf{P_{V\omega DE}}=\{\Omega_bh^2, \quad \Omega_ch^2, \quad 100\theta_{MC}, \quad \tau, \quad \omega, \quad \eta, \quad  \mathrm{ln}(10^{10}A_s), \quad  n_s \},   \label{12}
\end{equation}
\begin{equation}
\mathbf{P_{VKDE}}=\{\Omega_bh^2, \quad \Omega_ch^2, \quad 100\theta_{MC}, \quad \tau, \quad \Omega_{k}, \quad \eta, \quad  \mathrm{ln}(10^{10}A_s), \quad  n_s \},   \label{13}
\end{equation}
where $\Omega_bh^2$ and $\Omega_ch^2$ are the present-day baryon and CDM densities, $\theta_{MC}$ denotes the ratio between the angular diameter distance and the sound horizon at the redshift of last scattering $z^\star$, $\tau$ is the optical depth due to reionization, $\eta$ is bulk viscosity coefficient, $\omega$ is the perfect EoS of DE, $\Omega_{k}$ is the present-day spatial curvature,  $\mathrm{ln}(10^{10}A_s)$ and $n_s$ are the amplitude and spectral index of primordial power spectrum at the pivot scale $K_0=0.05$ Mpc$^{-1}$. Here $h$ is related to the Hubble constant $H_0$ by $h\equiv H_0/(100 \mathrm{kms^{-1}Mpc^{-1}})$.

\begin{figure}
\centering
\includegraphics[scale=0.36]{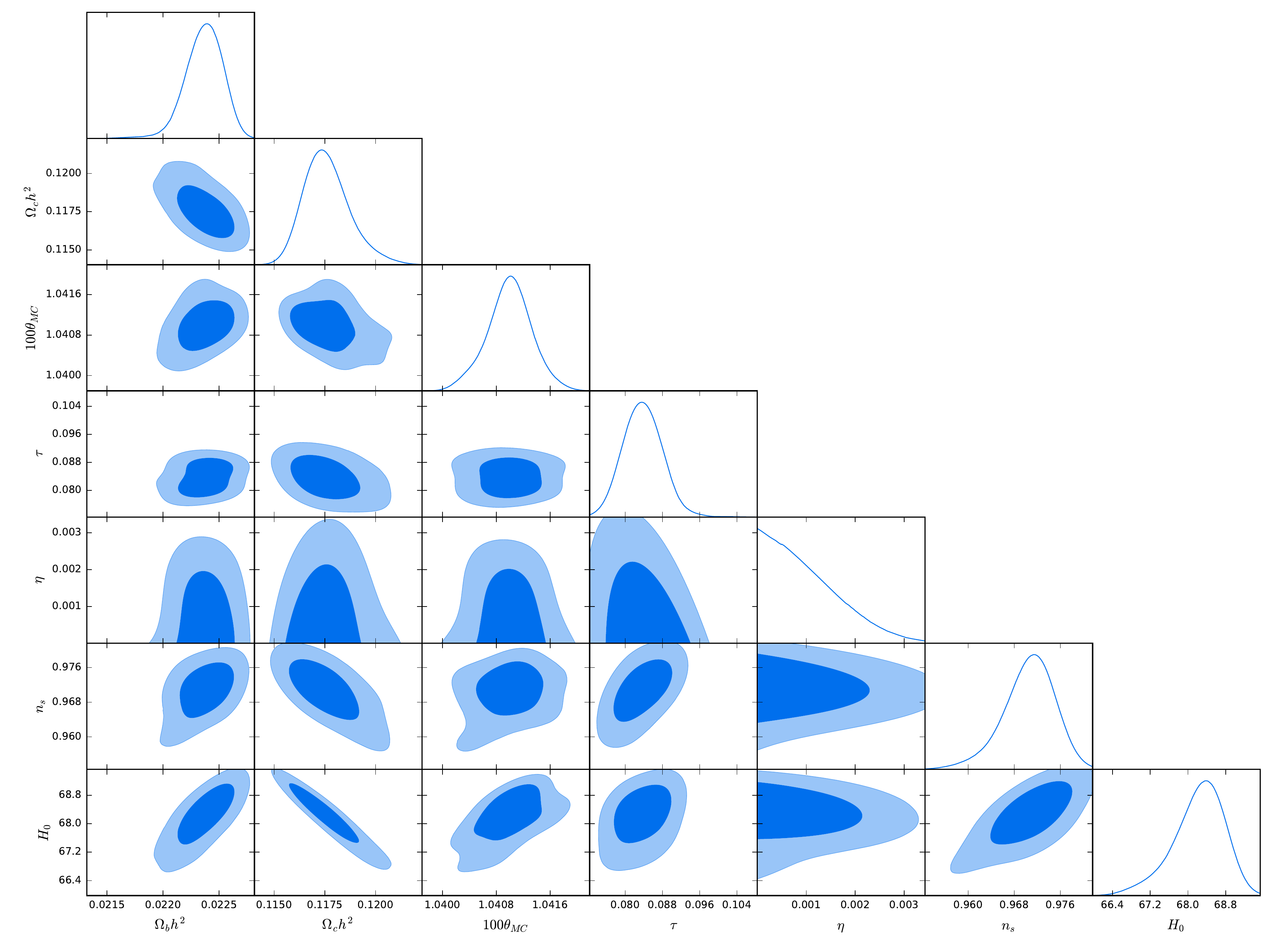}
\caption{The 1-dimensional posterior distributions on the individual parameters and $2$-dimensional marginalized contours of the V$\Lambda$DE model using the data combination CSBCHLW.}\label{f5}
\end{figure}

\begin{figure}
\centering
\includegraphics[scale=0.4]{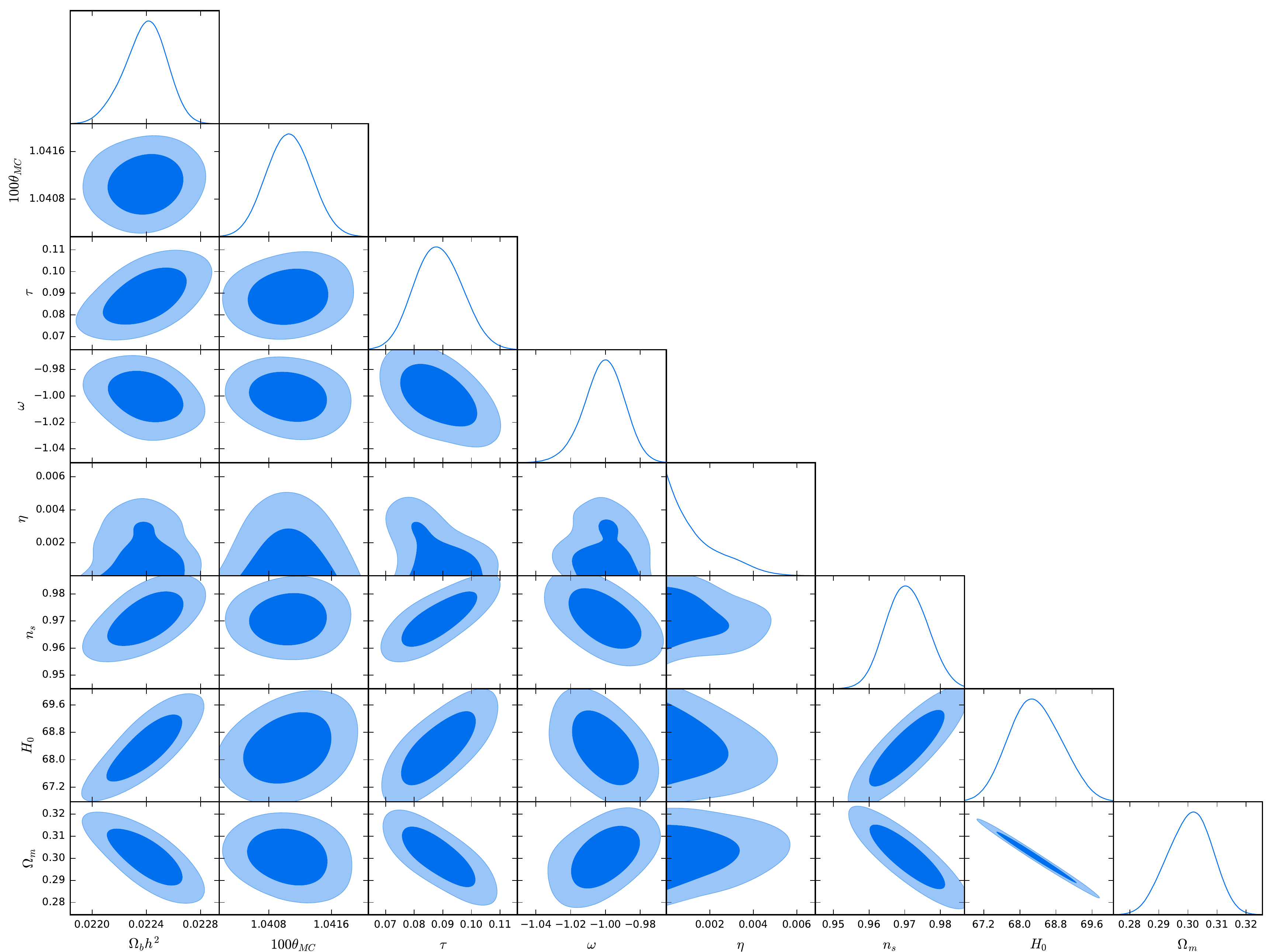}
\caption{The 1-dimensional posterior distributions on the individual parameters and $2$-dimensional marginalized contours of the V$\Lambda$DE model using the data combination CSBCHLW.}\label{f6}
\end{figure}

\begin{figure}
\centering
\includegraphics[scale=0.36]{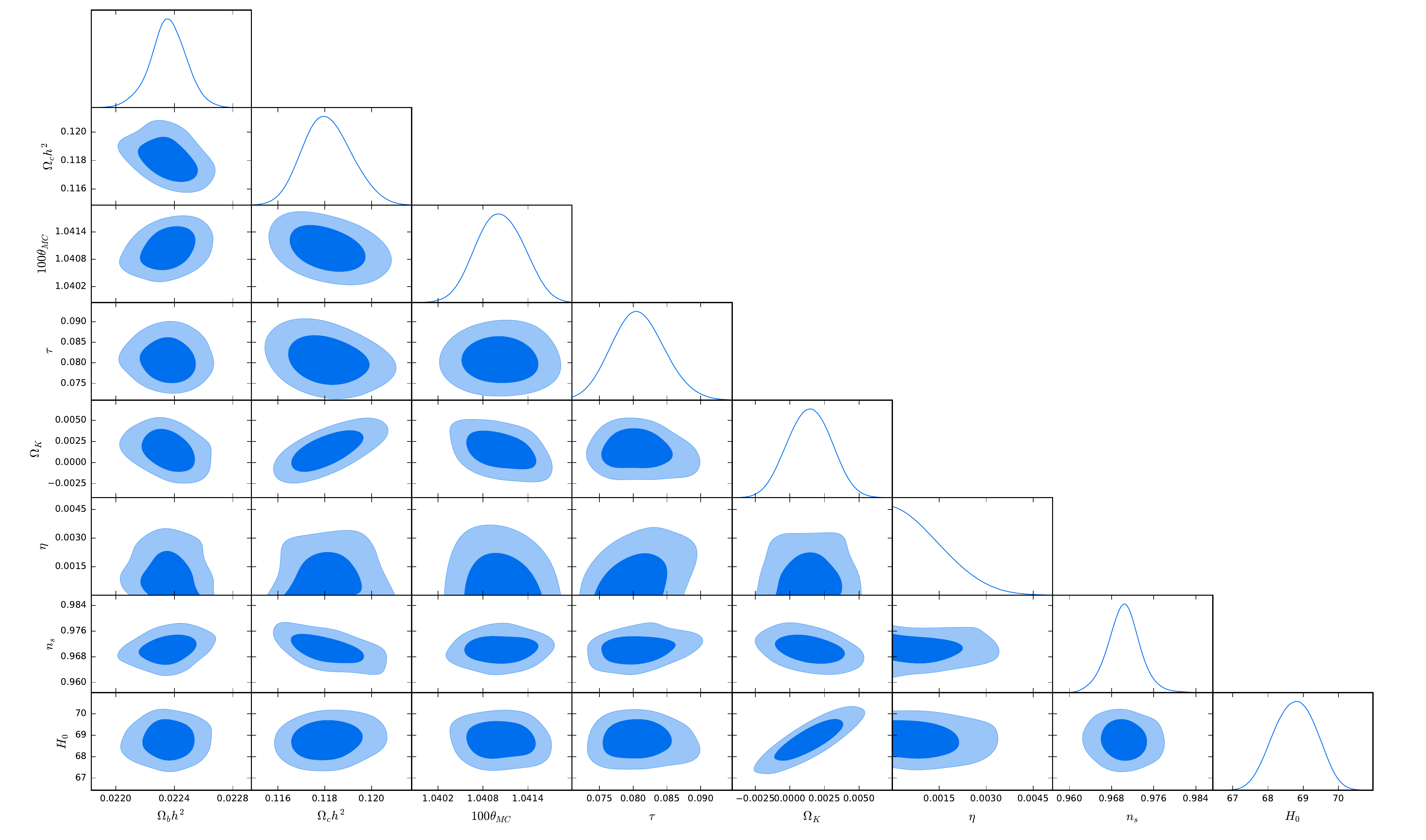}
\caption{The 1-dimensional posterior distributions on the individual parameters and $2$-dimensional marginalized contours of the V$\omega$DE model using the data combination CSBCHLW.}\label{f7}
\end{figure}

\begin{figure}
\centering
\includegraphics[scale=0.4]{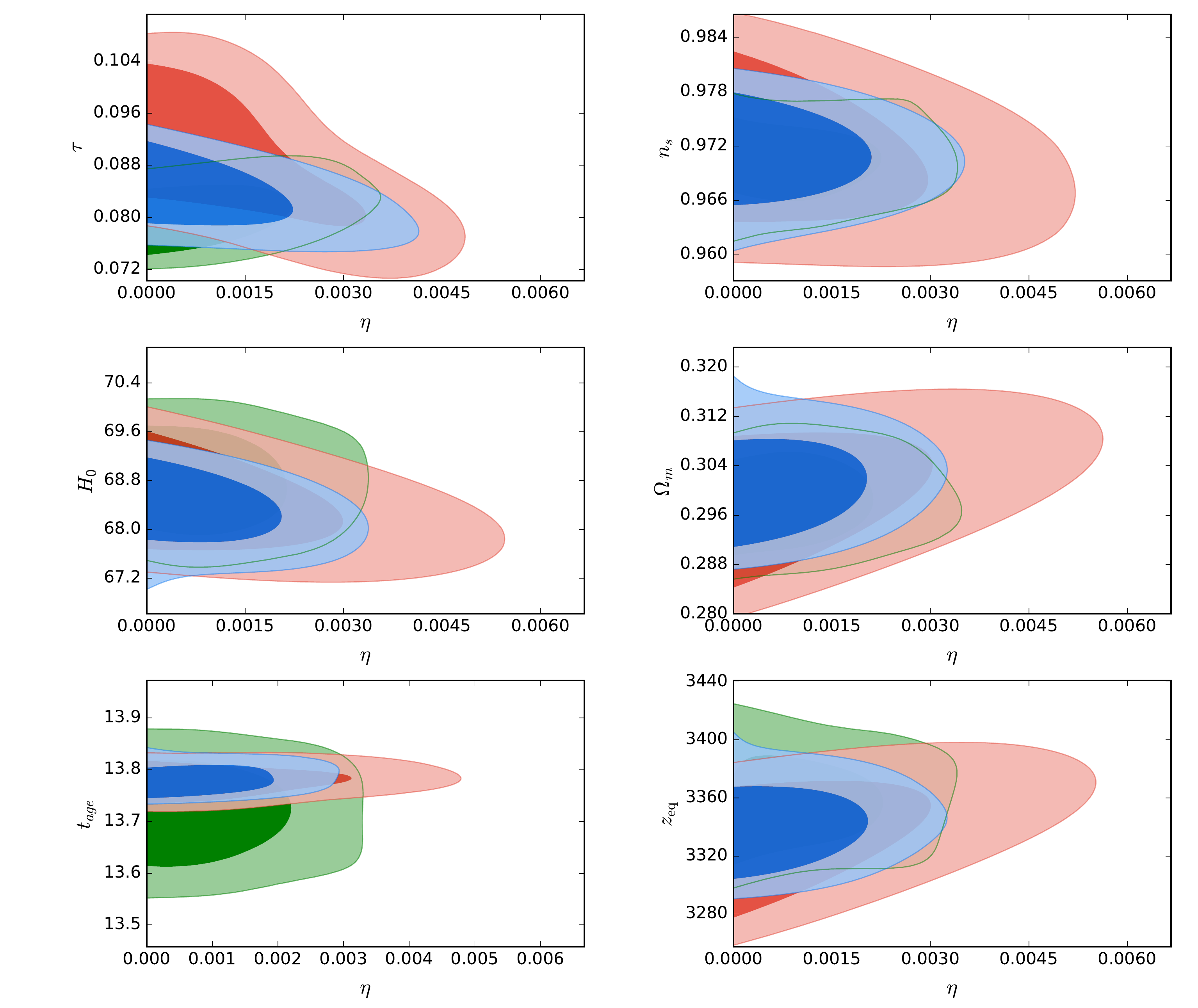}
\caption{The $2$-dimensional marginalized contours of the V$\Lambda$DE (blue), V$\omega$DE (red) and VKDE (green) models in the planes of $\eta-\tau$, $\eta-n_s$, $\eta-H_0$, $\eta-\Omega_m$, $\eta-t_{age}$ and $\eta-z_{eq}$ by using the data combination CSBCHLW, respectively.}\label{f8}
\end{figure}
\begin{figure}
\centering
\includegraphics[scale=0.4]{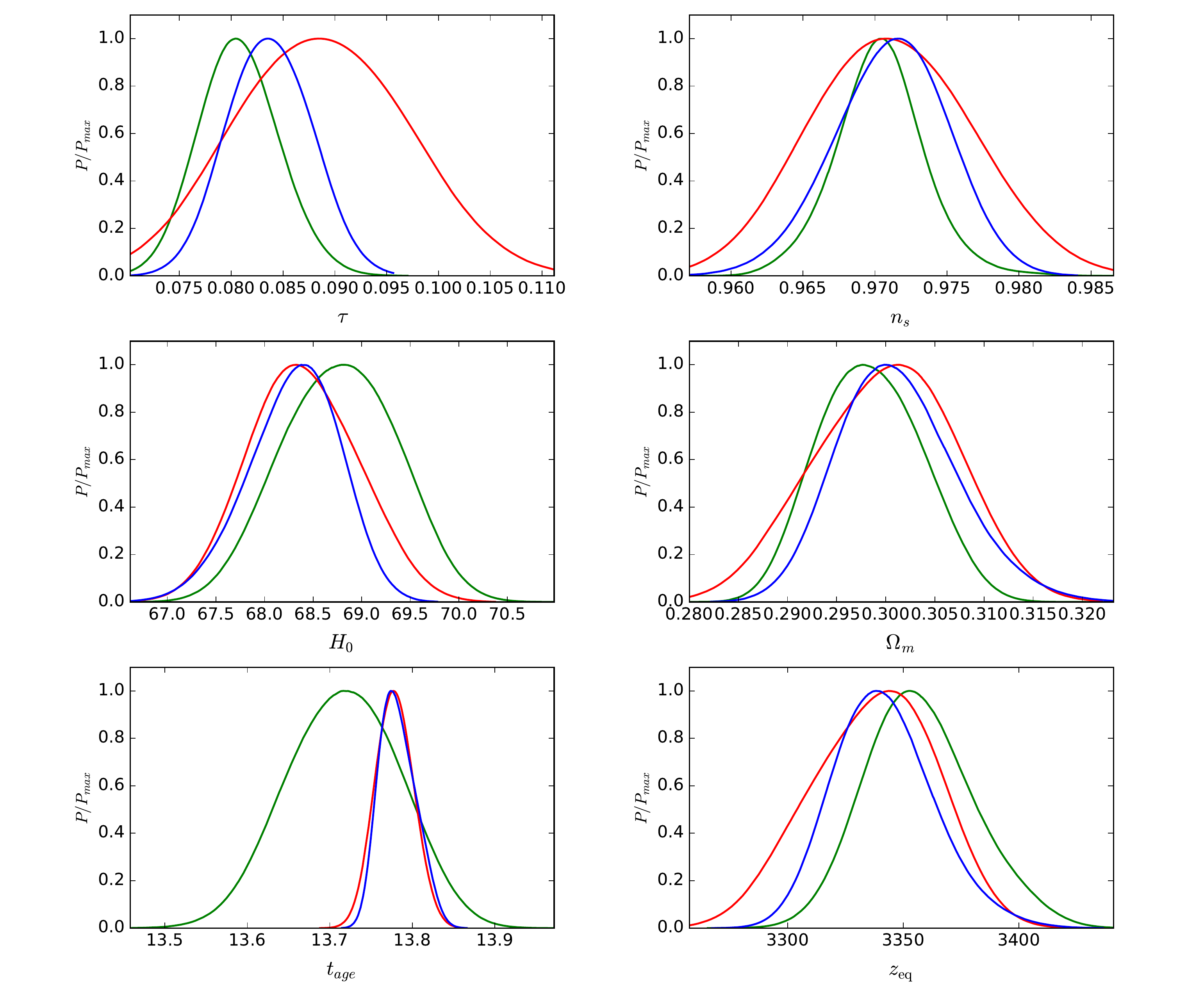}
\caption{The 1-dimensional posterior distributions of $\tau$, $n_s$, $H_0$, $\Omega_m$, $t_{age}$ and $z_{eq}$ in the V$\Lambda$DE (blue), V$\omega$DE (red) and VKDE (green) models using the data combination CSBCHLW, respectively.}\label{f9}
\end{figure}
\begin{figure}
\centering
\includegraphics[width=5.7cm,height=4.8cm]{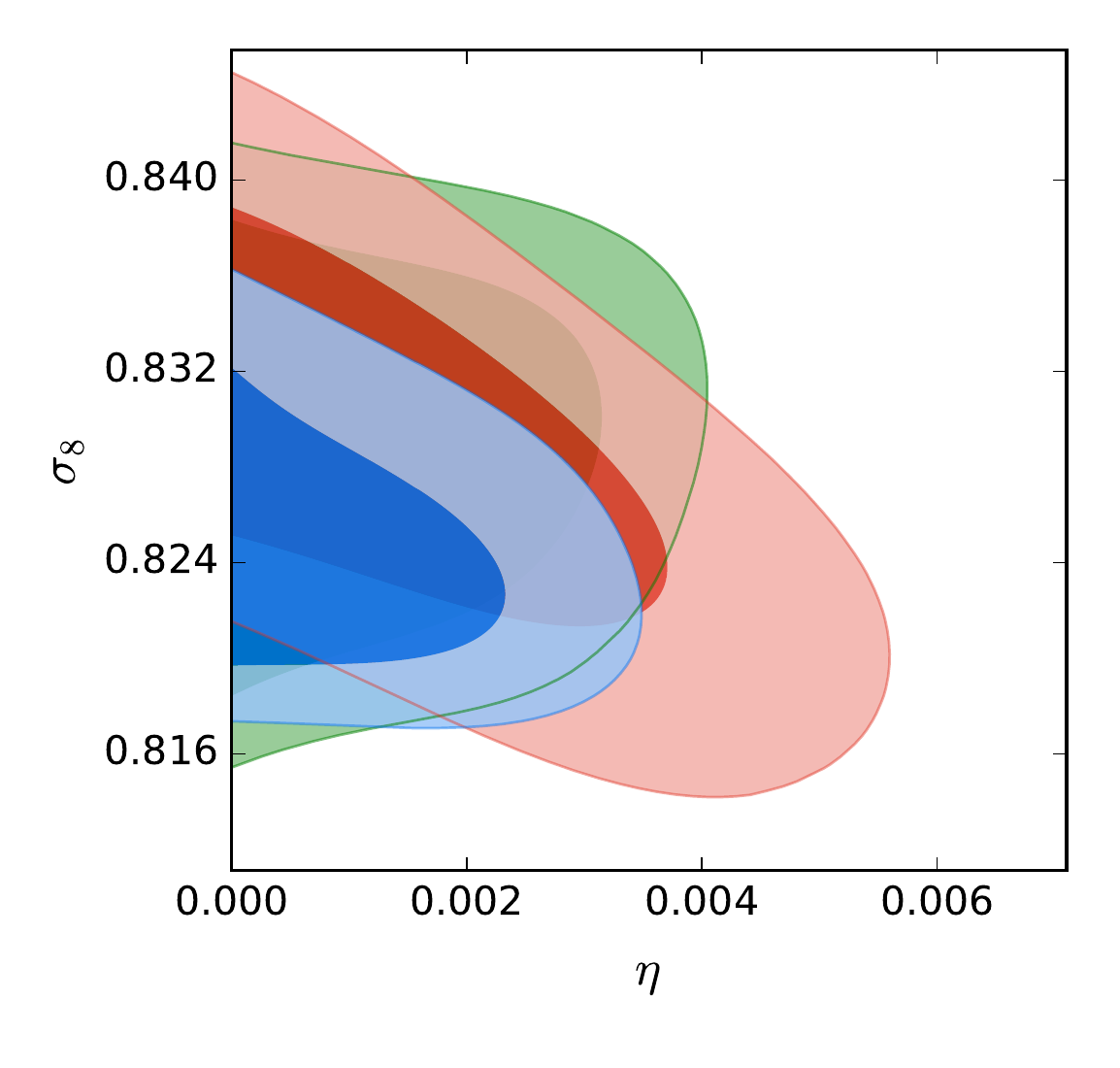}
\includegraphics[width=5.5cm,height=5cm]{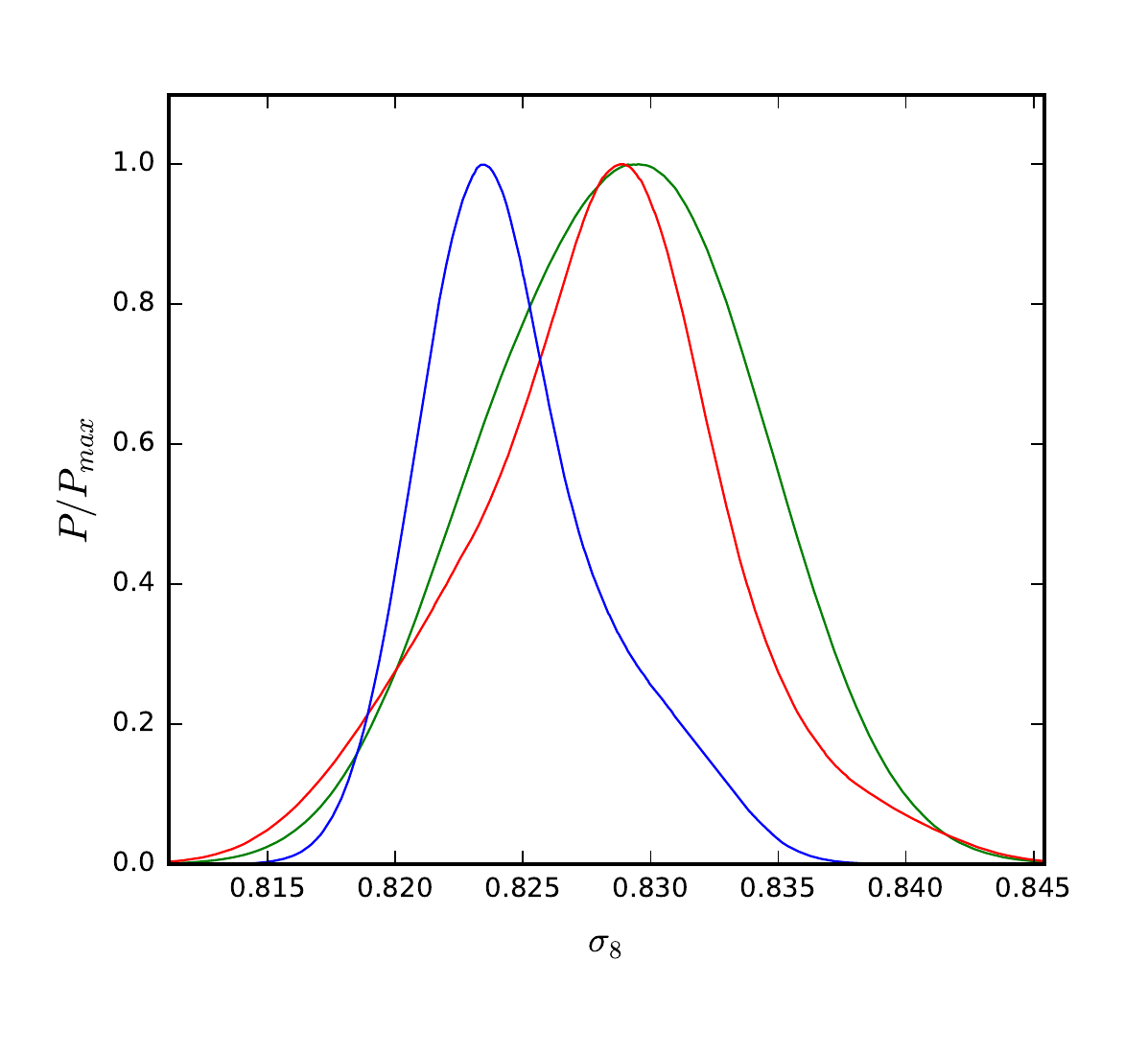}
\caption{The $2$-dimensional marginalized contour in the $\eta-\sigma_8$ plane and 1-dimensional posterior distribution of $\sigma_8$ for the V$\Lambda$DE (blue), V$\omega$DE (red) and VKDE (green) models using data combination CSBCHLW.}\label{f10}
\end{figure}

\section{Observations and methodology}
To study quantitatively the properties of bulk viscosity cosmology, we perform global constraints on our three VDE models by using the latest cosmological observations, which are exhibited as follows:

$\bullet$ \textbf{CMB}: The CMB data provides a substantially powerful way of constraining the background properties for a given cosmological model such as the matter constituents, topology and late-time evolution of the universe. In this analysis, we use the Planck 2015 temperature and polarization data \cite{6}, which includes the likelihoods of temperature (TT) at $30\leqslant \ell\leqslant 2500$, the cross-correlation of temperature and polarization (TE), the polarization (EE) power spectra, and the Planck low-$\ell$ temperature and polarization likelihood at $2\leqslant \ell\leqslant 29$.

$\bullet$ \textbf{SNIa}: SNIa is a powerful probe of cosmology, especially of EoS of DE. In 2013, Betoule et al. \cite{72} reported the results of an extensive campaign to improve the relative photometric calibrations between the SNLS and SDSS SN surveys, and constructed the largest SNIa sample `` Joint Light-curve Analysis '' (JLA) until now including SNLS, SDSS, HST and several samples of low-$z$ SN. This sample consists of 740 SNIa data points covering the redshift range $z\in[0.01, 1.3]$.

$\bullet$ \textbf{BAO}: BAO are also excellent geometrical probes and largely unaffected by errors in the nonlinear evolution of the matter density field and other systematic uncertainties which may have an impact on other cosmological data. To break the parameter degeneracies from other measurements, we employ four BAO data points: the 6dFGS sample at effective redshift $z_{eff}=0.106$ \cite{73}, the SDSS MGS sample at $z_{eff}=0.15$ \cite{74}, and the LOWZ at $z_{eff}=0.32$ and CMASS $z_{eff}=0.57$ samples of the SDSS-III BOSS DR12 sample \cite{75}.

$\bullet$ \textbf{Cosmic chronometer} (\textbf{CC}): The observational Hubble parameter data can be divided into two classes, i.e., the $H(z)$ data extracted from BAO surveys and CC data. In general, to obtain $H(z)$ data from BAO observations, one has to model the redshift space distortions and assume an acoustic scale, both of which require the assumption of a specific model. As a consequence, this class of data is model-dependent and cannot be used for constraining a cosmological model. Nonetheless, the CC data determined by using the most massive and passively evolving galaxies based on the `` galaxy differential age '' method are model-independent. We refer the readers to Ref. \cite{76} for detailed information about CC data (see also Ref. \cite{77}). Therefore, we use 30 CC data points in our analysis covering the range $z\in[0.07, 1.97]$.

$\bullet$ \textbf{Lensing} (\textbf{L}): Gravitational lensing by the large-scale structure leaves hints on the CMB temperature and polarization that can be measured in high angular resolution, low noise observations, such as those from Planck. To constrain the background parameters better, we also include the Planck lensing likelihood in our analysis, which has given the most powerful measurement with a 2.5$\%$ constraint on the amplitude of the lensing potential power spectrum \cite{78}.

Using the above cosmological observations, we adopt the Markov Chain Monte Carlo (MCMC) method to explore the parameter spaces of our three VDE models. We modifies the public package CosmoMC \cite{79} and Boltzmann code CAMB \cite{80} to infer the posterior probability distributions of different parameters. To perform the Bayesian analysis, we choose uniform priors for different parameters as shown in Tab. \ref{t1}. In succession, as studying the effects of different data combinations on the parameter estimations is beyond the scope of this study, we shall implement the tightest constraint on these VDE models using the data combination CMB + SNIa + BAO + CC + L, which is abbreviated as `` CSBCL '' in the following analysis.

\begin{table}[h!]
\renewcommand\arraystretch{1.3}

\caption{The prior ranges and the best-fit values and $1\sigma$ marginalized uncertainties of cosmological parameters in the V$\Lambda$DE, V$\omega$DE, and VKDE models using the data combination CSBCL, respectively. Note that we have quoted the (2$\sigma$) 95$\%$ limits for the parameters which cannot be well constrained.}
\label{t1}
\begin{tabular} { l l c c c }

\hline
\hline
Parameters & \quad Priors & V$\Lambda$DE & V$\omega$DE & VKDE\\
\hline
{$\Omega_b h^2   $} & $[0.005, 0.1]$       & $0.02237^{+0.00017}_{-0.00015}$ & $0.02239^{+0.00017}_{-0.00014}$ & $0.02233^{+0.00015}_{-0.00011}  $  \\

{$\Omega_c h^2   $} & $[0.001, 0.99]$      &
$0.11758^{+0.00093}_{-0.00120}        $ & $0.1174^{+0.0014}_{-0.0012}$ & $0.1183^{+0.0011}_{-0.0013}$ \\

{$100\theta_{MC} $} & $[0.5, 10]$    & $1.04099\pm 0.00033       $ & $1.04106\pm 0.00027 $  & $1.04101\pm 0.00030 $ \\

{$\tau           $} &$[0.01, 0.8]$ & $0.0837\pm 0.0037  $ & $0.0884\pm 0.0073   $ & $0.0807^{+0.0030}_{-0.0034}  $
                                                        \\

{${\rm{ln}}(10^{10} A_s)$} & $[2, 4]$    & $3.0945\pm 0.0071  $ & $3.104\pm 0.013    $  & $3.0974\pm 0.0075  $ \\

{$n_s            $} & $[0.8, 1.2]$        & $0.9707^{+0.0048}_{-0.0038}$ & $0.9706\pm 0.0053 $ & $0.9695^{+0.0040}_{-0.0029}$ \\

{$\eta         $} & $[0, 1]$ & $< 0.00217$ (2$\sigma$) & $< 0.00375 $  (2$\sigma$)  & $< 0.00282    $ (2$\sigma$)                                                  \\

{$\omega          $} & $[-2, 0]$ & --- &  $-1.001^{+0.012}_{-0.011}  $  & ---                                                    \\

{$\Omega_{k}          $} & $[-0.05, 0.05]$ & --- &  ---    & $0.0017\pm 0.0018   $                                                    \\

\hline
$H_0              $ & $[20, 100]$     & $68.24^{+0.61}_{-0.43}     $ & $68.34^{+0.55}_{-0.61}      $ & $68.81\pm 0.62      $                                                     \\

$\Omega_m              $ & \quad --- & $0.3020^{+0.0054}_{-0.0078}         $ & $0.3007^{+0.0080}_{-0.0071}   $ & $0.2985\pm 0.0054      $ \\

$\sigma_8              $ & \quad --- & $0.8252^{+0.0039}_{-0.0047}$ & $0.8285^{+0.0045}_{-0.0032}   $ & $0.8298\pm 0.0056    $ \\

\hline
$\chi^2_{min}$ & \quad --- & 13715.842 & 13714.371 & 13713.944 \\
\hline
\hline
\end{tabular}
\end{table}

\section{Analysis results}
Making the use of the tightest constraint we can give, the numerical analysis results are exhibited in Tab. \ref{t1} and Figs. \ref{f5}-\ref{f7}, which include the best-fit points and corresponding $1\sigma$ errors (or $2\sigma$ bounds) of individual parameters and 1-dimensional posterior distributions on the individual parameters and $2$-dimensional marginalized contours of three VDE models. It is easy to see that the V$\omega$DE and VKDE models give a similar $\chi^2_{min}$ and they give better cosmological fits than the V$\Lambda$DE one by the increasements of $\Delta\chi^2_{min}=1.471$ and 1.898, respectively. For three VDE models, we obtain, respectively, the $2\sigma$ upper bounds of the typical parameter $\eta$ 0.00217, 0.00375 and 0.00375 (see Tab. \ref{t1}). One can find that the latter two models give larger $2\sigma$ upper limits than the former one, which maybe ascribed to the enlargement of parameter spaces in the V$\omega$DE and VKDE models. For the V$\omega$DE model, we get the perfect EoS of DE $\omega=-1.001^{+0.012}_{-0.011}$, which is consistent with the prediction $\omega=-1.006\pm0.045$ of Planck under the assumption of $\omega$CDM at the $1\sigma$ confidence level (CL) \cite{6}. One can find that the perfect EoS of DE $\omega$ is still compatible with the standard cosmology ($\omega=-1$) and the best-fit value still prefers very slightly a phantom cosmology ($\omega<-1$), even if we consider the bulk viscosity effects of DE fluid. Furthermore, we also find that the spatial curvature $\Omega_{k}=0.0017\pm 0.0018$ of our universe in the VKDE model is consistent with zero at the $1\sigma$ CL, which implies that a spatially flat universe is supported by current data in the framework of viscous cosmology. Meanwhile, our result prefers a positive best-fit value corresponding to an open universe and is in good agreement with the restriction $|\Omega_{k}|<0.005$ of Planck. Additionally, we conclude that the scale invariant Harrison-Zeldovich-Peebles (HZP) power spectrum ($n_s=1$) \cite{81,82,83} is strongly excluded at the 6.10$\sigma$, 5.55$\sigma$ and 7.63$\sigma$ CL in the V$\Lambda$DE, V$\omega$DE and VKDE models, respectively, and that their predictive values of the spectral index $n_s$ are well consistent with the Planck evaluation $n_s=0.9655\pm0.0062$ at the $1\sigma$ CL \cite{6}. Recently, the improved local measurement $H_0=73.24\pm1.74$ km s$^{-1}$ Mpc$^{-1}$ from Riess et al. 2016 \cite{71} exhibits a strong tension with the Planck 2015 release $H_0=66.93\pm0.62$ km s$^{-1}$ Mpc$^{-1}$ \cite{70} at the $3.4\sigma$ CL. Using the data combination CSBCL,  we find that the current $H_0$ tension can be effectively alleviated from 3.4$\sigma$ to 2.71$\sigma$, 2.69$\sigma$ and 2.40$\sigma$ in the V$\Lambda$DE, V$\omega$DE and VKDE models, respectively, and that the consideration of curvature effect in the viscous cosmology appears to behave a little better than the other two models.

Utilizing the joint constraint CSBCL, we are also very interested in investigating the correlations between the typical parameter $\eta$ and other cosmological parameters in three VDE models. From Fig. \ref{f8}, we find that the present expansion rate $H_0$ is anti-correlated with the bulk viscosity coefficient $\eta$, which indicates that the universe expands slower with increasing viscosity of DE and is well consistent with the above qualitative analysis (see also Eq. (\ref{5}) and Fig. \ref{f1}). Subsequently, we conclude that the viscosity $\eta$ is positively correlated with the present matter density $\Omega_{m0}$ and the redshift of radiation-matter equality $z_{eq}$, which implies that the larger viscous effects of DE, the larger the matter fraction in the cosmic pie is and the later the time when the radiation and matter densities equal is. We also find that the optical depth $\tau$ is anti-correlated with the bulk viscosity $\eta$ in the V$\Lambda$DE and V$\omega$DE models, which means that the larger the viscosity of DE is, the smaller the optical depth is.  However, this is not the case in the VKDE one, where $\tau$ is positively correlated with $\eta$. Moreover, we find that the age of the universe increases very slightly with increasing bulk viscosity and the spectral index appears to decrease very slowly with increasing $\eta$.

To study the details of constrained parameters further and compare conveniently with each other in three VDE models, we exhibit their 1-dimensional posterior distributions in Fig. \ref{f9}. We find that the value of the optical depth of VKDE model is larger than those of the left two models (see Tab. \ref{t1}), and the values of the optical depth of these three VDE models are a little higher than the Planck result $\tau=0.066\pm0.012$ by using CMB and several low-$z$ probes \cite{6}. Meanwhile, we conclude that, since the values of $H_0$ and $t_{age}$ of VKDE model has evident shifts with respect to V$\Lambda$DE and V$\omega$DE ones, the VKDE model gives a larger expansion rate and a lower age of our universe than the other two ones (see Fig. \ref{f8}). Furthermore, three VDE scenarios give similar values of $n_s$, $\Omega_m$ and $z_{eq}$. In addition, from Fig. \ref{f10}, we find that the amplitude of the rms density fluctuations today $\sigma_8$ in linear regime is anti-correlated with the viscosity $\eta$, which indicates that the effect of matter clustering increases with increasing bulk viscosity of DE. One can also conclude that the V$\Lambda$DE model gives a lower value of the amplitude of matter fluctuations than the left two scenarios from the right panel of Fig. \ref{f10} (see Tab. \ref{t1}).

\section{Discussions and conclusions}
Our motivation is to study the effects of bulk viscosity on the evolution of the universe using the latest cosmological observations. Specifically, we propose three new VDE models, i.e., V$\Lambda$DE, V$\omega$DE, and VKDE, investigate their background evolutional behaviors via qualitatively numerical analysis, and place constraints on them using the data combination CSBCL. For three VDE models, we obtain the corresponding $2\sigma$ upper limits of the bulk viscosity coefficient. We find that the V$\omega$DE and VKDE models give a similar $\chi^2_{min}$, while they provide better cosmological fits than the V$\Lambda$DE one, that these three VDE models can alleviate effectively the current $H_0$ tension from 3.4$\sigma$ to 2.71$\sigma$, 2.69$\sigma$ and 2.40$\sigma$, respectively, and that the consideration of curvature effect in the VKDE model appears to behave a little better in reliving the $H_0$ tension than the other two models. For the V$\omega$DE and VKDE models, we find that our restrictions on the prefect EoS of DE and the spatial curvature are well consistent with the predictions by the Planck experiment.

Furthermore, for these three VDE models, we also obtain the following conclusions: (i) The scale invariance of HZP primordial power spectrum is strongly excluded, while their constrained values of spectral index are in good agreement with the Planck result \cite{6}; (ii) The universe expands slower with increasing bulk viscosity; (iii) With increasing bulk viscosity, the age of the universe increases very slightly and the spectral index appears to decrease very slowly; (iv) The effect of matter clustering increases with increasing bulk viscosity of DE.

The addition of curvature in the VDE scenario can lead to some interesting conclusions: (i) The VKDE model can relieve the $H_0$ tension a little better than the other two ones; (ii) The optical depth of VKDE model is anti-correlated with the bulk viscosity, which is not the case in the left two scenarios; (ii)The VKDE model gives a larger value of optical depth, a lower evolutional age and a larger expansion rate of the universe. These conclusions may imply that there exist deeper correlations between the cosmic curvature and basic cosmological quantities. Undoubtedly, the spatial curvature is a direct probe of the spacetime topology, and the detection of a significant deviation from $\Omega_k=0$ would have profound meanings for inflationary cosmology and fundamental physics. This topic is worth being discussed further in the future work.

Since our results give very tight restrictions on the bulk viscosity coefficient $\eta$ in three VDE models, combining with constraints on the perfect EoS of DE and the spatial curvature, one can easily find that the VDE models just deviate very slightly from the standard cosmology based on the current cosmological data (see also Figs. \ref{1}-\ref{4}). We expect that future high-precision gravitational wave observations can give new information about the evolution of the universe.

\section{acknowledgements}
We thank B. Ratra and S. D. Odintsov for helpful discussions on cosmology. D. Wang is grateful to L. Xu, Y. Li, W. Yang, F. Melia and Y. Sun for useful  communications. This work is partly supported by the National Science Foundation of China.

\end{document}